\documentstyle[12pt]{article}
\begin{document}

\title{ On The Ladder Bethe-Salpeter Equation}
\author{G. V. Efimov \vspace*{0.2\baselineskip}\\
 \itshape Bogoliubov Laboratory of Theoretical Physics,\\
 \itshape Joint Institute for Nuclear Research, Dubna,
 Russia\vspace*{0.2\baselineskip} }
%
% \date{Pacs Numbers: 13.20.-v, 13.20.Fc, 13.20.He, 24.85.+p}
%
\maketitle
%..........
\begin{abstract}

The Bethe-Salpeter (BS) equation in the ladder approximation is
studied within a scalar theory: two scalar fields (constituents)
with mass $m$ interacting via an exchange of a scalar field (tieon)
with mass $\mu$. The BS equation is written
in the form of an integral equation in the configuration
Euclidean $x$-space with the kernel which for stable bound states $M<2m$
is a self-adjoint positive operator. The solution of the BS equation
is formulated as a variational problem. The nonrelativistic
limit of the BS equation is considered. The role of so-called
abnormal states is discussed.

The analytical form of test functions for which the accuracy of
calculations of bound state masses is better than 1\% (the comparison
with available numerical calculations is done) is determined.
These test functions make it possible to calculate analytically
vertex functions describing the interaction of bound states with
constituents.

As a by-product a simple solution of the Wick-Cutkosky model
for the case of massless bound states is demonstrated.

\end{abstract}
%..............................
%\pacs{Pacs Numbers: 13.20.-v, 13.20.Fc, 13.20.He, 24.85.+p}
%..............................

\newpage

\section{Introduction}

The ladder Bethe-Salpeter equation (BS) is an effective if not unique
instrument to study bound states within the framework of quantum field
theory (see, for example, \cite{Wick,Itz,Sapir,Grein,Nak}). However,
the exact analytical solution is obtained only for the Wick-Cutkosky
model in a particular case when the masses of intermediate particle
and bound state are equal to zero \cite{Itz,Nak}. For other
cases analytical solutions are not found and numerical methods are
applied. Numerous literature is devoted to numerical solutions
of the ladder Bethe-Salpeter equation for different particles,
propagators and interactions to get acceptable description of
spectrum and other characteristics in nuclear and particle physics.
Although computer numerical methods are very powerful,
we are convinced that a simple method to get analytical solutions
of the Bethe-Salpeter equation in the form of known special functions
with accuracy around 0.1-1.0\% would be very useful for qualitative
and semiquantative analysis of physical phenomena, the more if
the accuracy can be improved. At present, the accuracy around 1\%
is undoubtedly acceptable in particle phenomenology at low energies.

A generally accepted approach to solve the BS equation
( I do not know exceptions ) consists in investigation of this equation
in the momentum space where the BS-amplitude is expanded on a
suitable basis in the Euclidean apace ${\bf R}^4$ to reduce the original
four-dimensional integral equation to an infinite set of coupled
one-dimensional equations which should be solved by numerical methods
(see, for example, \cite{Itz,Tjon,Kapt} and many-many other papers).

The goal of this work is to improve the variational approach to the
bound-state problem of two scalar particles with mass $m$ interacting
through the exchange of a scalar particle with mass $\mu$ within the
ladder Bethe-Salpeter equation. In addition, variational calculations
are attractive because they give restrictions on computable parameters
from one definite side, meanwhile approximate calculations do not
indicate what side we approach to an exact value from. Variational
methods were applied to the BS equation long time ago \cite{Vosko,Schw}
and now they are used directly in the Hamiltonian formulation of
quantum field theory (see, for example, \cite{Dar}).
Our idea is the following. In the configuration Euclidean $x$-space
the BS equation can be rewritten as an integral equation with the kernel
which for stable bound states $M<2m$ is a self-adjoint positive symmetric
operator. Thus, the problem of solution is reduced to searching
eigenvalues and eigenfunctions of this kernel. As a result,
the Bethe-Salpeter amplitude is expanded over this orthonormal
system of functions and this series is nothing but the representation
of the BS-amplitude over the Regge poles in the $s$-channel. The same
representation of the BS-amplitude can be written in the $t$-channel.
In this approach, the explicit form and normalization of vertex functions
is naturally determined by the eigenfunctions. Besides, this
representation makes clear the role of so-called {\it abnormal states}.
Abnormal states are part of the full orthonormal system of functions.
They are needed to get the correct representation of the Bethe-Salpeter
amplitude over bound states. The experimental status of abnormal states
is not clear up to now. I think this problem deserves special
investigation.

Since the kernel is a self-adjoint positive and symmetric operator,
the task to find eigenvalues and eigenfunctions can be formulated as
a variational problem. It turned out that quite simple
test functions describe the bound state masses with accuracy
less than 1\% (we compare our results with the numbers obtained
in \cite{Tjon}).

It is interesting to discuss the connection between the Schr\"{o}dinger
and BS equations and to estimate qualitatively the value of the coupling
constant for which relativistic corrections should be taken into account.
It turned out that the nonrelativistic limit is realized for very
small coupling constants $\alpha\leq 10^{-3}$. In other words, any
situations when binding energy is larger than 1\% of constituent
masses, require relativistic description.

\section{The Bethe-Salpeter equation in the self-adjoint form
and its formal solution}

Within a simple quantum field model we shall consider the
simplest case of the Yukawa interaction of charged scalar particles
("constituents") described by the field $\Phi$ with mass $m$
and neutral intermediate bosons ("tieons") described by the field $\phi$
with mass $\mu$. The Lagrangian density is
\begin{eqnarray}
\label{lagr}
L(x)&=&\Phi^+(\Box-m^2)\Phi+{1\over2}\phi(\Box-\mu^2)\phi+g\Phi^+\Phi\phi.
\end{eqnarray}
For simplicity, in the following we put $m=1$. It means that all
dimensional variables ($x$-coordinate, energy, momentum and masses)
are expressed in units of the constituent mass $m$.
This model is frequently used as the simplest pattern of QFT in
many discussions, although this system is considered to be unstable
from a strict point of view because the Hamiltonian is not bounded from
below (see, for example, \cite{Simon}). Nevertheless this model
has been investigated by various methods (see, for example,
\cite{Itz,Wick,Bog,Dar} and references therein).

A two-particle bound state corresponds to a solution of the homogeneous
Bethe-Salpeter equation for a four-point amplitude. This equation
in the ladder approximation in the configuration Euclidean $x$-space
looks like
\begin{eqnarray}
\label{BSx}
&&\Box(x_1,x_2;x_3,x_4)=D(x_1-x_2)\delta(x_1-x_3)\delta(x_2-x_4)\\
&&+g^2\int\!\!\!\int dz_1dz_2D(x_1-x_2)S(x_1-z_1)S(x_2-z_2)
\Box(z_1,z_2;x_3,x_4),\nonumber
\end{eqnarray}
where the propagators of constituents and tieons are
\begin{eqnarray}
\label{SD}
&& S(x)=\int{dp\over(2\pi)^4}\cdot{e^{ipx}\over 1+k^2}=
{1\over(2\pi)^2}{1\over\sqrt{x^2}}{\rm K}_1(\sqrt{x^2}),\\
&& D(x)=\int{dk\over(2\pi)^4}\cdot{e^{ikx}\over \mu^2+k^2}=
{1\over(2\pi)^2}{\mu\over\sqrt{x^2}}{\rm K}_1(\mu\sqrt{x^2}).\nonumber
\end{eqnarray}
Let us write down equation (\ref{BSx}) in the center-of-mass frame,
i.e. let us introduce the variables
$$x_1=x+{y\over2},~~x_2=x-{y\over2},~~x_3=x'+{y'\over2},~~
x_4=x'-{y'\over2},~~z_1=x''+{y''\over2},~~~z_2=x''-{y''\over2}.$$
Then
\begin{eqnarray*}
&&\Box(x_1,x_2;x_3,x_4)=\Box(x-x';y,y')
\end{eqnarray*}
and equation (\ref{BSx}) reads
\begin{eqnarray}
\label{BSxx}
\Box(x-x';y,y')&=&\sqrt{D(y)}\delta(x-x')\delta(y-y')\sqrt{D(y')}\\
&+&g^2\int\!\!\!\int dx''dy''D(y)\Pi(x-x'',y-y'')\Box(x''-x';y'',y'),
\nonumber
\end{eqnarray}
\begin{eqnarray*}
&& \Pi(x-x',y-y')=
S\left(x-x'+{y-y'\over2}\right)S\left(x-x'-{y-y'\over2}\right)
\end{eqnarray*}
Now let us go to the Fourier transform
\begin{eqnarray*}
&& \Box(x-x';y,y')=\int{dp\over(2\pi)^4}e^{ip(x-x')}\Box_p(y,y')
\end{eqnarray*}
and define the function
\begin{eqnarray*}
&& \Box_p(y,y')=\sqrt{D(y)}Z_p(y,y')\sqrt{D(y')}
\end{eqnarray*}
Then equation (\ref{BSxx}) becomes
\begin{eqnarray}
\label{Zyy}
&& Z_p(y,y')=\delta(y-y')+g^2\int dy''K_p(y,y'')Z_p(y'',y'),
\end{eqnarray}
with the symmetric kernel
\begin{eqnarray}
\label{Kp}
&& K_p(y,y')=\sqrt{D(y)}\Pi_p(y-y')\sqrt{D(y')}
\end{eqnarray}
\begin{eqnarray}
\label{Pi}
\Pi_p(y-y')&=&\int{dx\over(2\pi)^4}e^{-ipx}
S\left(x+{y-y'\over2}\right)S\left(x-{y-y'\over2}\right)\\
&=&\int{dk\over(2\pi)^4}{e^{ik(y-y')}\over
\left(\left(k+{p\over2}\right)^2+1\right)
\left(\left(k-{p\over2}\right)^2+1\right)}\nonumber\\
&=&\int{dk\over(2\pi)^4}{e^{ik(y-y')}\over
\left(k^2+1+{p^2\over4}\right)^2-(kp)^2}.\nonumber
\end{eqnarray}

We want to emphasize that equation (\ref{Zyy}) represents
the Bethe-Salpeter equation in the form of the integral
equation with the symmetrical kernel $K_p(y,y')$.
If we consider the problem of stable bound states, i.e. $p^2=-s=-M^2$
and $M<2$ in (\ref{Pi}), then the kernel $K_p(y,y')$ is the
self-adjoint positive symmetrical normal operator on the functional space
${\bf L}^2$. This kernel can be represented in the form
\begin{eqnarray}
\label{KpU}
&& K_p(y,y')=\sum\limits_QU_Q(y,p)\Lambda_Q(s)U_Q(y,p),~~~~~(s=-p^2)
\end{eqnarray}
where the functions $\Lambda_Q(s)$ and $U_Q(y,p)$ are the
eigenvalues and eigenfunctions of the kernel $K_p(y,y')$, i.e.,
\begin{eqnarray}
\label{LU}
&& \Lambda_Q(s)U_Q(y,p)=\int dy'K_p(y,y')U_Q(y',p).
\end{eqnarray}
Below we discuss the quantum numbers $Q$.

The functions $\{U_Q(y,p)\}$ are orthonormal
\begin{eqnarray*}
&& \int dy~U_Q(y,p)U_{Q'}(y,p)=\delta_{QQ'},~~~~~
\sum\limits_QU_Q(y,p)U_Q(y',p)=\delta(y-y')
\end{eqnarray*}

The general properties of the eigenvalues $\Lambda_Q$ are
characterized by the traces
\begin{eqnarray}
\label{trace}
&& T_n={\rm Tr}~K_p^n=\sum\limits_Q\Lambda_Q^n(s)=\int\!\!...\!\!
\int dy_1...dy_n~K_p(y_1,y_2)...K_p(y_n,y_1).
\end{eqnarray}
The trace (\ref{trace}) for the kernel (\ref{Kp}) is finite
$T_n<\infty$ for $n\geq3$ in the case $\mu=0$ and for $n\geq2$
in the case $\mu>0$.
It means that eigenvalues decrease as
$\Lambda_Q(s)\sim{1\over |Q|^c}$ with an appropriate constant $c$.

If the eigenfunctions and eigenvalues are found, one can write
\begin{eqnarray*}
&& \delta(y-y')+g^2K_p(y,y')=\sum\limits_Q
U_Q(y,p)[1+g^2\Lambda_Q(s)]U_Q(y',p).
\end{eqnarray*}
Let us represent
\begin{eqnarray*}
&& Z_p(y,y')=\sum\limits_Q U_Q(y,p)\tilde{Z}_Q(p)U_Q(y',p),
\end{eqnarray*}
then equation (\ref{LU}) looks like
\begin{eqnarray*}
&& \tilde{Z}_Q(p)=1+g^2\Lambda_Q(s)\tilde{Z}_Q(p),~~~~
 \tilde{Z}_Q(p)={1\over1-g^2\Lambda_Q(s)}.
\end{eqnarray*}
One gets
\begin{eqnarray*}
&& Z_p(y,y')=\sum\limits_Q U_Q(y,p){1\over1-g^2\Lambda_Q(s)}U_Q(y',p).
\end{eqnarray*}
and
\begin{eqnarray}
\label{Box}
&& \Box_p(y,y')=\sum\limits_Q
\sqrt{D(y)}U_Q(y,p){1\over1-g^2\Lambda_Q(s)}U_Q(y',p)\sqrt{D(y')}
\end{eqnarray}

Finally, the solution of the Bethe-Salpeter equation (\ref{BSx})
in the $x$-space is
\begin{eqnarray}
\label{BSsol}
&&\Box(x_1,x_2;x_3,x_4)=\Box(x-x';y,y')\\
&&=\sum\limits_Q\int{dp\over(2\pi)^4}e^{-ipx}
\sqrt{D(y)}U_Q(y,p){1\over1-g^2\Lambda_Q(s)}U_Q(y',p)\sqrt{D(y')}e^{ipx'}.
\nonumber
\end{eqnarray}

The standard Fourier transform of the amplitude $\Box$ in the momentum space
looks like
\begin{eqnarray*}
&&\int...\int \!dx_1dx_2dx_3dx_4~e^{i(k_1x_1+k_2x_2-k_3x_3-k_4x_4)}
\Box(x_1,x_2;x_3,x_4)\\
&&=(2\pi)^4\delta(k_1+k_2-k_3-k_4)\Box(s,t).
\end{eqnarray*}
Here $s=-(k_1+k_2)^2,~~t=-(k_1-k_3)^2$, where $k_1$, $k_2$ and $k_3$
are the Euclidean momenta. Thus, the amplitude $\Box$ as the
solution of the Bethe-Salpeter equation (\ref{BSx}) reads
\begin{eqnarray}
\label{Bst}
&& \Box(s,t)=\sum\limits_QV_Q(q,p){1\over1-g^2\Lambda_Q(s)}V_Q(q',p)
\end{eqnarray}
where $ p=k_1+k_2,~~~~q=k_1-k_2,~~~~q'=k_3-k_4$ and the vertices are
\begin{eqnarray}
\label{VQ}
&& V_Q(q,p)=i^l\int dy~e^{-iqy}\sqrt{D(y)}U_Q(y,p)
\end{eqnarray}

The mass of a bound state with quantum numbers $Q$ is
defined by the equation
\begin{eqnarray}
\label{mass}
&& 1=g^2\Lambda_Q(M_Q^2),~~~~~~~~ p=({\bf 0},iM_Q) .
\end{eqnarray}

Formula (\ref{Bst}) gives the Regge representation over the poles
in the $s$-channel. The same representation is valid in the $t$-channel.

\section{Variational solution of the Bethe-Salpeter equation}

We start with the following observation. The usual variation approach
is applied to the BS equation written in the well-known standard form
(see, for example, \cite{Vosko,Schw})
\begin{eqnarray}
\label{EqL}
&& L\psi=\lambda D\psi,
\end{eqnarray}
where $L$ and $D$ are self-adjoint positive operators. The result is
\begin{eqnarray}
\label{VarL}
&& \lambda=\min\limits_\psi{(\psi L\psi)\over(\psi D\psi)}.
\end{eqnarray}
For a test function $\Psi$ we get
\begin{eqnarray}
\label{lambda0}
&& \lambda\leq\lambda_0={(\Psi L\Psi)\over(\Psi D\Psi)}.
\end{eqnarray}
However, if we rewrite equation (\ref{EqL}) in the form
\begin{eqnarray}
\label{EqLD}
&& {1\over \lambda}\phi=
\sqrt{D}{1\over L}\sqrt{D}\cdot \phi,~~~~~~\phi=\sqrt{D}\psi,
\end{eqnarray}
the variational problem looks like
\begin{eqnarray}
\label{VarLD}
&& {1\over \lambda}=\max\limits_\phi
{\left(\phi\sqrt{D}{1\over L}\sqrt{D}\phi\right)\over(\phi \phi)}=
\max\limits_\psi{\left(\psi D{1\over L}D\psi\right)\over(\psi D\psi)}
\end{eqnarray}
and for the same test function $\Psi$ as in (\ref{lambda0}) we get
\begin{eqnarray}
\label{lambda1}
&& {1\over \lambda}\geq{1\over\lambda_1}=
{\left(\Psi D{1\over L}D\Psi\right)\over(\Psi D\Psi)},~~~~
\lambda\leq\lambda_1
\end{eqnarray}
One of the corollaries of the Cauchy inequality for a positive
operator $L$ is
\begin{eqnarray*}
&& (X~Y)^2\leq(X~L~X)\left(Y{1\over L}Y\right),~~~~~\forall X,Y\in{\bf L}^2.
\end{eqnarray*}
This inequality for $X=\Psi$ and $Y=D\Psi$ gives
\begin{eqnarray}
\label{lambda01}
&& {(\Psi D\Psi)\over\left(\Psi D{1\over L}D\Psi\right)}
\leq{\left(\Psi L\Psi\right)\over(\Psi D\Psi)}~~~~{\rm and}~~~~
\lambda_1\leq\lambda_0
\end{eqnarray}
Thus, the variational method is more preferable to apply to
equation (\ref{EqLD}) and use (\ref{VarLD}).

Let us come back to equation (\ref{LU}). The kernel $K_p(y,y')$
in (\ref{Kp}) for stable states $p^2=-s=-M^2,~~~M<2m$ is the
self-adjoint positive operator. The spectrum of this operator
is bounded from above. The main preference of the representation
(\ref{Kp}) is that variational methods can be used to find the
spectrum and eigenfunctions of this kernel. Therefore, the solution of
equation (\ref{LU}) can be reduced to the variational problem
in the form (\ref{VarLD})
\begin{eqnarray}
\label{VPr}
&& \Lambda_Q(s)=\max\limits_{U_Q}{(U_Q K_pU_Q)\over(U_QU_Q)}
=\max\limits_{U_Q}{(U_Q\sqrt{D}\Pi_p\sqrt{D}U_Q)\over(U_QU_Q)}
\end{eqnarray}
It is convenient to introduce the function
\begin{eqnarray}
\label{Psi}
&& U_Q(y,p)=\sqrt{D(y)}\Psi_Q(y,p).
\end{eqnarray}
Then formula (\ref{VPr}) reads
\begin{eqnarray}
\label{VPsi}
&& \Lambda_Q(s)=\max\limits_{\Psi_Q}{(\Psi_QD\Pi_pD\Psi_Q)\over
(\Psi_QD\Psi_Q)}
\end{eqnarray}
where
\begin{eqnarray*}
&& (\Psi_QD\Pi_pD\Psi_Q)=\int\!\!\!\int\! dydy'~\Psi_Q(y,p)D(y)
\Pi_p(y-y')D(y')\Psi_Q(y',p),\\
&& (\Psi_QD\Psi_Q)=\int\! dy~\Psi_Q(y,p)D(y)\Psi_Q(y,p).
\end{eqnarray*}
The vertex (\ref{VQ}) is defined by
\begin{eqnarray}
\label{VVQ}
&&V_Q(q,p)=i^l\int dy~e^{-iqy}D(y)\Psi_Q(y,p).
\end{eqnarray}
The representation (\ref{VPsi}) will be exploited in this paper.

Another possibility is to introduce the function
\begin{eqnarray}
\label{Psi}
&& U_Q(y,p)={\Phi_Q(y,p)\over\sqrt{D(y)}}.
\end{eqnarray}
Then formula (\ref{VPr}) becomes
\begin{eqnarray}
\label{VPhi}
&& \Lambda_Q(s)=\max\limits_{\Phi_Q}{(\Phi_Q\Pi_p\Phi_Q)\over
\left(\Phi_Q{1\over D}\Phi_Q\right)}
\end{eqnarray}
This representation is particularly convenient in the case
of the Wick-Cutkosky model ($\mu=0$) for which
\begin{eqnarray}
\label{Dgl}
&& D(y)={1\over(2\pi)^2}\cdot{1\over y^2}~~~~{\rm and}~~~~
{1\over D(y)}=(2\pi)^2\cdot y^2.
\end{eqnarray}

The variational problem in the form (\ref{VPsi}) and (\ref{VPhi}) is
the key object of our investigation.

The mass of a bound state with quantum numbers $Q$ is
defined by equation (\ref{mass}). The eigenvalue $\Lambda_Q(M)$
is a monotonically increasing function when the mass $M$ increases
on the interval $M^2\in[0,4m^2]$, i.e.
$$ \Lambda_Q(M_1^2)<\Lambda_Q(M_2^2)~~~{\rm if}~~~M_1<M_2.$$
For a test function $\Psi$ in (\ref{VPsi}) we get
\begin{eqnarray}
\label{inequ}
&& \Lambda_Q(M^2)\geq\Lambda_Q^{(var)}(M^2).
\end{eqnarray}
It gives
\begin{eqnarray*}
&& {1\over g^2}=\Lambda_Q(M_Q^2)=\Lambda_Q^{(var)}((M_Q^{(var)})^2)
\geq\Lambda_Q^{(var)}(M_Q^2).
\end{eqnarray*}
Thus, the variational calculations in (\ref{VPr},\ref{VPsi},\ref{VPhi})
give the upper estimation for bound state masses
\begin{eqnarray}
\label{mvar}
&& M_Q\leq M_Q^{(var)}.
\end{eqnarray}

If we want to find the coupling constant for a given mass $M$, then
\begin{eqnarray*}
&& {1\over g_{(var)}^2}=\Lambda_Q^{(var)}(M_Q^2)\leq
\Lambda_Q(M_Q^2)={1\over g^2}
\end{eqnarray*}
and
\begin{eqnarray}
\label{mvar}
&& g^2\leq g^2_{(var)}.
\end{eqnarray}

\subsection{Classification of states}

Let us consider the classification of states dictated by equation
(\ref{LU}). The kernel $K_p(y,y')$ depends on the direction
of the vector $p$. It is convenient to work in the frame
$p=({\bf 0},p_4)=({\bf 0},iM)$.

Let us start with the case $M=0$, i.e. $p=0$. Then the ladder
Bethe-Salpeter equation has $O(4)$ symmetry in the Euclidean
configuration space and the eigenfunctions
look like
\begin{eqnarray}
\label{eigenf}
&& \Psi_{n\kappa\{\mu\}}(y)=T^{(\kappa)}_{\mu_1...\mu_\kappa}(y)
R_{n\kappa}(y^2),\\
&& T^{(\kappa)}_{\{\mu\}}(y)=O(|y|^\kappa),~~~
R_{n\kappa}(y^2)=O(1)~~~{\rm for}~~~y\to0,\nonumber
\end{eqnarray}
where $n$ and $\kappa$ are radial and orbital quantum numbers and
$T^{(\kappa)}_{\mu_1...\mu_\kappa}(y)$ are spherical harmonics in
the space ${\bf R}^4$ (see Appendix). All states are degenerated
over four-dimensional orbital quantum number $\kappa$ and the number of
these states is $(\kappa+1)^2$ for fixed $n$ and $\kappa$.

Now let $M>0$. Then the Bethe-Salpeter equation possesses $O(3)$
symmetry and all states are degenerated over three-dimensional
orbital quantum numbers $l$ which define the spin of corresponding
bound states. The degeneration of the state (\ref{eigenf}) is taken off
and the total number of these states can be represented
$$(\kappa+1)^2=\sum\limits_{l=0}^\kappa(2l+1),$$
i.e. after removal of degeneration, the state (\ref{eigenf}) turns into
superposition of states with three-dimensional orbital momenta
$l\leq \kappa$. Usually the state with $l=\kappa$ is considered as the
desired bound state with spin $l$, but the states with $l<\kappa$,
describing the spin $l$, are called "abnormal" ones and are not considered.
From the point of wiew of the representation (\ref{eigenf}) the number
$\kappa$ characterizes the behaviour of the wave function in the vicinity
of zero, i.e. $\Psi_{n\kappa l\{\mu\}}\sim|y|^\kappa$ for $y\to0$.
Unconditionally the parameter $\kappa$ cannot be a normal quantum number,
but for numerical calculations with a fixed accuracy  $\kappa$ is
connected with the smallness parameter because in real calculations
the additional factor $|y|^\kappa$ leads to decreasing of integrals.
These argumentations are consistent with the standard approach
(see, for example, \cite{Tjon}), when the BS amplitude is developed
over the four-dimensional hyperspherical harmonics
$Z_{\kappa l m}(\chi,\theta,\phi)$ and the obtained system of equations
is truncated over a "nonphysical" quantum number $\kappa$.

Thus, we can characterize a bound state by the quantum numbers
$Q=(n,\kappa,l,m)$ where $(l,m)$ are three-dimensional orbital quantum
numbers, $n$ is a radial quantum number and $\kappa$ is connected with
the behaviour of a wave function for $|y|\to0$. The eigenvalues are
degenerated over the magnetic quantum number $m$, i.e.
$$\Lambda_Q(s)=\Lambda_{n\kappa l}(s).$$

On the other hand, one can say that the quantum number $\kappa$ is
connected with "time" or the fourth coordinate component $y_4$.
It has no special name and usually these states are called
"abnormal". These states disappear in the nonrelativistic limit
and usually are not considered. However, the representation of the
Bethe-Salpeter amplitude (\ref{Bst}) is not valid without these
states and their role should be clarified.

We choose the test functions $\Psi_Q$ in the form
\begin{eqnarray}
\label{test}
\Psi_Q(y,p)&=&\Psi_{n\kappa lm}(y,p)=
Y_{lm}({\bf y})W_{n\kappa l}(|{\bf y}|,y_4)\\
&=&Y_{lm}({\bf y})R_{n\kappa l}(|{\bf y}|,y_4)F_{n\kappa l}(|{\bf y}|,y_4).
\nonumber
\end{eqnarray}
Here $Y_{lm}({\bf y})=Y_{lm}({\bf n}_y)|{\bf y}|^l$ with
${\bf n}_y={\bf y}/|{\bf y}|$. $Y_{lm}({\bf n}_y)=Y_{lm}(\theta,\phi)$
is the standard three-dimensional spherical function;
$R_{n\kappa l}(u,v)$ is a polynomial of the degree $(n+\kappa)$ and
$F_{n\kappa l}(u,v)$ is a test function. The constant $N_{n\kappa l}$
is defined by the normalization condition
\begin{eqnarray}
\label{norm}
&& (\Psi_QD\Psi_{Q'})=\int\!\! dyD(y)\Psi_{n\kappa lm}(y,p)
\Psi_{n'\kappa'l'm'}(y,p)=
\delta_{ll'}\delta_{mm'}\delta_{nn'}\delta_{\kappa\kappa'}
\end{eqnarray}
The vertex (\ref{VQ}) takes the form
\begin{eqnarray}
\label{vertex}
V_Q(q,p)&=&V_{n\kappa lm}(q,p)=i^l\int dy~e^{-iqy}D(y)
Y_{lm}({\bf y})W_{n\kappa l}(|{\bf y}|,y_4)\\
&=&Y_{lm}({\bf q})V_{n\kappa l}(q,p).\nonumber
\end{eqnarray}
where
\begin{eqnarray*}
&& V_{n\kappa l}(q,p)=
\int\!\!\!\int d{\bf y}dy_4~|{\bf q}|^l|{\bf y}|^l
P_l({\bf n}_q{\bf n}_y)e^{-i({\bf qy}+q_4y_4)}
D\left(\sqrt{{\bf y}^2+y_4^2}\right)W_{n\kappa l}(|{\bf y}|,y_4)
\end{eqnarray*}
Finally, the BS amplitude (\ref{Bst}) looks like
\begin{eqnarray}
\label{Bst-nl}
&& \Box(s,t)=\sum\limits_{n\kappa l}{2l+1\over4\pi}
V_{n\kappa l}(q,p){1\over1-g^2\Lambda_{n\kappa l}(s)}V_{n\kappa l}(q',p)
\end{eqnarray}

The test function $F(u,v)$ contains variational parameters. We choose
the test function in the following simplest form
\begin{eqnarray}
\label{testWD}
&& W(y)=W(|{\bf y}|,|y_4|)=e^{-a\sqrt{{\bf y}^2+by_4^2}},\\
&& D_3(y)=D_3(|{\bf y}|,|y_4|)
=\int{dk\over(2\pi)^4}\cdot{e^{iky}\over(a^2+{\bf k}^2+bk_4^2)^3},\nonumber
\end{eqnarray}
where $a$ and $b$ are variational parameters.

Polynomials $R_{n\kappa l}(u,v)$ have to provide the orthogonality condition
(\ref{norm}).

\section{Nonrelativistic limit}

\subsection{\bf The "Bethe-Salpeter form" of the nonrelativistic
Schr\"{o}dinger equation }

First of all let us represent the nonrelativistic Schr\"{o}dinger equation
in the "Bethe-Salpeter form". Bound states of two particles with
masses $m=1$ interacting via an attractive potential are described
by the equation
\begin{eqnarray}
\label{Sch}
&& H\psi({\bf x})=\left({\bf p}^2-\alpha V(r)\right)\psi({\bf x})=
-\epsilon\psi({\bf x}).
\end{eqnarray}
where $\alpha$ is the coupling constant, $\epsilon$ is the binding energy
and
$${\bf x}\in{\bf R}^3,~~r=\sqrt{{\bf x}^2},~~{\bf p}^2=-{\bf \partial_x}^2.$$
The function $V(r)>0$ is supposed to be positive. For the Yukawa
potential $V(r)={e^{-\mu r}\over r}$.

Let us rewrite equation (\ref{Sch}) in the form
\begin{eqnarray*}
&&\left({\bf p}^2+\epsilon\right)\psi({\bf x})=
\alpha V(r)\psi({\bf x}),~~~~~
\psi({\bf x})=\alpha{1\over{\bf p}^2+\epsilon}
\cdot V(r)\cdot\psi({\bf x})
\end{eqnarray*}
and let us introduce the function
\begin{eqnarray*}
&&\phi({\bf x})=\sqrt{V(r)}\psi({\bf x})
\end{eqnarray*}
Then we get
\begin{eqnarray*}
\phi({\bf x})&=&\alpha\sqrt{V(r)}\cdot{1\over{\bf p}^2+\epsilon}
\cdot\sqrt{V(r)}\phi({\bf x})=
\alpha \int d{\bf x}'~S_\epsilon({\bf x},{\bf x}')\phi({\bf x}').
\end{eqnarray*}
Here the kernel $S_\epsilon({\bf x},{\bf x}')$ for $\epsilon>0$ is the
self-adjoint symmetric positive operator
\begin{eqnarray}
\label{S}
S_\epsilon({\bf x},{\bf x}')&=&\sqrt{V(r)}\cdot
{1\over-{\bf \partial_x}^2+\epsilon}\delta({\bf x}-{\bf x}')
\cdot\sqrt{V(r')}\nonumber\\
&=&\sqrt{V(r)}\cdot
\Pi_\epsilon({\bf x}-{\bf x}')\cdot\sqrt{V(r')},\nonumber\\
\Pi_\epsilon({\bf x-x'})&=&\int{d{\bf k}\over(2\pi)^3}
{e^{-i{\bf k(x-x')}}\over{\bf k}^2+\epsilon}={1\over4\pi}\cdot
{e^{-\sqrt{\epsilon}|{\bf x-x'}|}\over|{\bf x-x'}|}.
\end{eqnarray}
The solution of the Schr\"{o}dinger equation (\ref{Sch}) is equivalent
to the variational problem which can be written in different forms
\begin{eqnarray}
\label{varnBS}
&&1=\alpha\Lambda(\epsilon)=
\alpha\max\limits_\phi{(\phi S_\epsilon\phi)\over(\phi\phi)}=
\alpha\max\limits_\psi {(\psi V\Pi_\epsilon V\psi)\over(\psi V\psi)}=
\alpha\max\limits_\varphi {(\varphi \Pi_\epsilon \varphi)\over
\left(\varphi{1\over V}\varphi\right)}.\nonumber\\
\end{eqnarray}
We will show that in nonrelativistic limit $M\to2m$ the variational
problem (\ref{VPsi}) is reduced to (\ref{varnBS}).

\subsection{\bf The nonrelativistic limit
in the Bethe-Salpeter equation}

Now let us come back to the Bethe-Salpeter equation in the form
(\ref{VPhi}) and consider the polarization operator $\Pi_p(y-y')$
\begin{eqnarray*}
\Pi_p(y-y')&=&\int{d{\bf k}\over(2\pi)^3}\int{dk_4\over2\pi}
{e^{-i{\bf k}({\bf y}-{\bf y}')-ik_4(y_4-y_4')}\over
\left({\bf k}^2+k_4^2+1-{M^2\over4}\right)^2+M^2k_4^2}
\end{eqnarray*}
where $ p=({\bf 0},iM)$ and $(pk)=iMk_4 $.

For a state $\Psi(y,p)$ we have
and
\begin{eqnarray*}
(\Psi D\Pi_pD\Psi)&=&\int{d{\bf k}\over(2\pi)^3}\int{dk_4\over2\pi}\cdot
{\tilde{\Psi}^2({\bf k},k_4) \over
\left({\bf k}^2+k_4^2+\Delta\right)^2+M^2 k_4^2}.
\end{eqnarray*}
Here $\Delta=1-{M^2\over4}$ and
$$\tilde{\Psi}({\bf k},k_4)=\int dy~e^{ipy}D(y)\Psi({\bf y},y_4).$$

In the nonrelativistic limit
\begin{eqnarray}
\label{nonrel}
&& M\to2,~~~~~\Delta=1-{M^2\over4}\to 0.
\end{eqnarray}
Introducing new variables ${\bf k}=\sqrt{\Delta}~{\bf q}$ and
$k_4=\Delta q_4$ one can get in the limit (\ref{nonrel})
\begin{eqnarray*}
(\Psi D\Pi_PD\Psi)&=&\sqrt{\Delta}\int{d{\bf q}\over(2\pi)^3}
\int{dq_4\over2\pi}\cdot{\tilde{\Psi}^2(\sqrt{\Delta}{\bf q},\Delta q_4)
\over({\bf q}^2+\Delta q_4^2+1)^2+4q_4^2}\\
&\to&\sqrt{\Delta}\int{d{\bf q}\over(2\pi)^3}
\int{dq_4\over2\pi}{\tilde{\Psi}^2(\sqrt{\Delta}{\bf q},0)
\over({\bf q}^2+1)^2+4q_4^2}\\
&=&{1\over4}\sqrt{\Delta}\int{d{\bf q}\over(2\pi)^3}
{\tilde{\Psi}^2(\sqrt{\Delta}{\bf q},0)\over{\bf q}^2+1}
={1\over4}\int{d{\bf k}\over(2\pi)^3} {\Psi^2({\bf k},0)
\over{\bf k}^2+\Delta}
\end{eqnarray*}
The test function is chosen in the form
$\Psi(y,p)=\Psi({\bf y},0)=\psi({\bf y})$.
It leads to
\begin{eqnarray*}
\tilde{R}({\bf p},0)&=&\int d{\bf y}\int dy_4~
e^{-i{\bf py}}D(y)\psi({\bf y})=
{1\over4\pi}\int d{\bf y}{e^{-\mu|{\bf y}|}\over |{\bf y}|}
e^{-i{\bf py}}\psi({\bf y})\\
&=&{1\over4\pi}\int d{\bf y}e^{-i{\bf py}}V(|{\bf y}|)\psi({\bf y})
\end{eqnarray*}
and
\begin{eqnarray*}
&&(\Psi D\Psi)=\int\!\!\!\int d{\bf y}dy_4D(y)\psi^2({\bf y})
={1\over4\pi}\int d{\bf y}{e^{-\mu|{\bf y}|}\over|{\bf y}|}\psi^2({\bf y})
={1\over4\pi}\left(\psi V\psi\right)
\end{eqnarray*}
The binding energy equals
\begin{eqnarray*}
&& \Delta=(2-M)\cdot\left(1-{2-M\over4}\right)
=\epsilon\left[1-{\epsilon\over4}\right]\approx \epsilon
\end{eqnarray*}
Collecting all these formulas we obtain (\ref{varnBS})
\begin{eqnarray*}
&&1=\lambda(\epsilon)= \alpha\max\limits_\psi
{\left(\psi V\Pi_\epsilon V\psi\right)
\over\left(\psi V\psi\right)}
\end{eqnarray*}
where $\alpha$ is connected with the coupling constant $g$ in the
following way:
\begin{eqnarray}
\label{a-g}
&& \alpha={g^2\over16\pi}={g^2\over16\pi m^2}={\lambda\over4},~~~~~
 \lambda={g^2\over4\pi m^2}.
\end{eqnarray}

The main result for the solution of the variational problem
of the relativistic Bethe-Salpeter equation (\ref{VPsi})
for small binding energy $\epsilon\ll1$ is that it is convenient
to choose the test function $\Psi$ in the form
\begin{eqnarray}
\label{test}
&& \tilde{\Psi}(p)=\tilde{\Psi}(\sqrt{\Delta} a{\bf p},\Delta b p_4)
\end{eqnarray}
where $a$ and $b$ are variational parameters.

\section{The Wick-Cutkosky model $\mu=0$}

\subsection{The mass $M_Q=0$ in the Wick-Cutkosky model}

As a by-product of the representation (\ref{KpU}) equation (\ref{LU})
can be easily solved in the case $M_Q=0$ (see \cite{Wick,Nak}).
Equation (\ref{LU}) reads
\begin{eqnarray*}
&& U=g^2\Lambda U=g^2KU
\end{eqnarray*}
or
\begin{eqnarray}
\label{eqU}
U(y)&=&{g^2\over(2\pi)^2}{1\over\sqrt{y^2}}\int dy'\Pi_0(y-y')
{1\over\sqrt{(y')^2}}U(y')
\end{eqnarray}
We look for the solution in the form $U(y)=\sqrt{y^2}~\Phi(y')$.
Equation (\ref{eqU}) reads
$$ y^2\Phi(y)={g^2\over(2\pi)^2}\int dy'\Pi_0(y-y')\Phi(y')$$
and in the momentum space
\begin{eqnarray}
\label{eqPhi}
&& -\Box_k\tilde{\Phi}(k)={4\alpha\over\pi}\cdot{1\over(k^2+1)^2}
\tilde{\Phi}(k),~~~~~~~\alpha={g^2\over16\pi}.
\end{eqnarray}
The states are degenerated over quantum numbers $m$ and $\kappa=l$.
Let us extract the angular variables
\begin{eqnarray*}
&& \tilde{\Phi}(k)=T^{(l)}_{\mu_1...\mu_l}(k)A(k^2)
\end{eqnarray*}
where $T^{(l)}_{\mu_1...\mu_l}(k)$ is the spherical function in the
four-dimension space (see Appendix) and satisfies the correlations
(\ref{Tl})
$$\Box_kT^{(l)}_{\mu_1...\mu_l}(k)=0,~~~~
\left(k{\partial\over\partial k}\right)T^{(l)}_{\mu_1...\mu_l}(k)
=lT^{(l)}_{\mu_1...\mu_l}(k).$$
Then equation (\ref{eqPhi}) reads
with $k^2=u\in[0,\infty]$
\begin{eqnarray*}
&& u~A''(u)+(2+l)~A'(u)+{\alpha\over\pi}{1\over(u+1)^2}~A(u)=0.
\end{eqnarray*}
Let us introduce new variables
$$ t={u-1\over u+1}\in[-1,1],~~~~~~A(u)=(1-t)^{l+1}B(t). $$
We get
\begin{eqnarray}
\label{eqB}
&& (1-t^2)~B''(t)-2(2+l)~tB'(t)+\left({\alpha\over\pi}-l^2-3l-2\right)~B(t)=0
\end{eqnarray}
The solution is the Jacobi polynomial
$B(t)=P_n^{(l+1,l+1)}(t)$ with
$$ {\alpha\over\pi}-l^2-3l-2=n(n+2l+3)$$
and finally
\begin{eqnarray}
\label{spectr}
&& \alpha_{nl}=\pi(n+l+1)(n+l+2)
\end{eqnarray}

\subsection{Variation calculations}

For the Wick-Cutcosky model we perform variational calculations
for the lowest and first excited states and compare our numerical
results with the numbers obtained in \cite{Tjon}.

\subsubsection{The lowest state}

For the lowest state the test function is chosen in the form
\begin{eqnarray}
\label{psivar}
&& \Psi_0(y,p)=W(|{\bf y}|,|y_4|)=e^{-a\sqrt{{\bf y}^2+by_4^3}},
\end{eqnarray}
where $a$ and $b$ are variational parameters. Then
\begin{eqnarray}
\label{R}
V_0(k,p)&=&\int dy~D(y)W(|{\bf y}|,|y_4|)e^{-iky}=
\int{dy\over(2\pi)^2y^2}e^{-a\sqrt{{\bf y}^2+by_4^2}-i{\bf ky}-ik_4y_4}
\nonumber\\
&=&{ab+\sqrt{a^2b+ub+v}\over[u+v+a^2(1+b)+2a\sqrt{a^2b+ub+v}]
~\sqrt{a^2b+ub+v}}\nonumber\\
&=&R(u,v;a,b),~~~~~u={\bf k}^2,~~~v=k_4^2;
\end{eqnarray}
and
\begin{eqnarray*}
(\Psi D\Psi)&=&\int{dy\over(2\pi)^2y^2}e^{-2a\sqrt{{\bf y}^2+by_4^2}}
={1\over4a^2(1+\sqrt{b})}
\end{eqnarray*}
The eigenvalue $\Lambda(M)$ is defined by
\begin{eqnarray}
\label{varmass}
&& g^2\Lambda(M^2)=\alpha\max\limits_{a,b}{8\over\pi^2}
\int\limits_0^\infty dk~k^2\int\limits_0^\infty dv\cdot
{4a^2(1+\sqrt{b})R^2(k^2,v^2;a,b)\over
(k^2+v^2+\Delta)^2+4(1-\Delta)v^2}
\end{eqnarray}
The value of $M$ is defined by equation (\ref{mass}).
It is instructive to compare the Bethe-Salpeter mass (\ref{varmass})
with the bound state mass of the nonrelativistic Schr\"{o}dinger
equation for the Coulomb potential
\begin{eqnarray*}
&& \epsilon=2-M={\alpha^2\over4},~~~~~~~
{1\over\alpha}={1\over\sqrt{8\left(1-{M\over2}\right)}}=
{1\over\alpha_{Sch}},\\
&& \epsilon_{em}={\alpha_{em}^2\over4}=1.33...\cdot10^{-5},~~~~~
\alpha_{em}={1\over137}
\end{eqnarray*}
The numerical results are represented in Table 1 where we have used
$a=\sqrt{\Delta}a_1$ and $b=\Delta b_1$.

\begin{center}
{Table 1. The lowest state in the Wick-Cutkosky model. The coupling
constant $\lambda=4\alpha$ for different masses $M/m$ and comparison
$1/\alpha$ with the nonrelativistic value $1/\alpha_{Sch}$.}
\end{center}
\begin{center}
\begin{tabular}{|c||c|c||c||c|c||c|c|}
\hline
& & & & & & & \\
${M\over m}$&${\epsilon\over2m}$&${1\over\alpha_{Sch}}$&${1\over\alpha}$&
$a_1$&$b_1$&$\lambda=4\alpha$&$\lambda_{Tj}$\cite{Tjon} \\
& & & & & & &\\
\hline
& & & & & & &\\
0.000 & 1.00 & 0.3535  & 0.1586 & 0.76 & 1.0 & 25.21 & 25.13\\
0.500 & 0.75 & 0.4082  & 0.1671 & 0.78 & 1.0 & 23.93 &  \\
1.000 & 0.50 & 0.5000  & 0.1994 & 0.80 & 1.0 & 20.06 & 20.01\\
1.500 & 0.25 & 0.7071  & 0.3023 & 0.84 & 1.3 & 13.23 & \\
1.900 & 0.05 & 1.5811  & 0.8901 & 0.88 & 2.1 & 4.493 & 4.483 \\
1.999 & 0.0005 & 15.811&13.993  & 1.00 & 4.5 & 0.285 & 0.285\\
$2-{\alpha_{em}^2\over4}$ & ${\alpha_{em}^2\over8}$
& 137 &133.9 & 1.0 & 6.0 & 0.0299 &\\
& & & & & & & \\
\hline
\end{tabular}
\end{center}
One can see that the variational calculations practically coincide with
the numerical results obtained in \cite{Tjon}.

It is important to remark that the nonrelativistic regime, when
the Schr\"{o}dinger equation can be used, is realized for very
small binding energies $\epsilon\leq 10^{-4}$.

\subsubsection{The first orbital exited states}

The test functions for the "normal" state with $\kappa=0,~l=1$ and
for "abnormal" state with $\kappa=1,~l=0$ look like
\begin{eqnarray}
\label{sl11}
&& {\bf \Psi}^{(n)}(y,p)={\bf y}~W(|{\bf y}|,|y_4|),~~~~~~
\Psi^{(a)}(y,p)=y_4~W(|{\bf y}|,|y_4|)
\end{eqnarray}
where the test function $W$ is the first function in (\ref{testWD}).
One can get after simple calculations
\begin{eqnarray*}
{\bf V}^{(n)}(k,p)&=&i\int dy~{\bf y}D(y)W(|{\bf y}|,|y_4|)e^{-iky}\\
&=&-2{\bf k}{\partial\over\partial u}R(u,v;a,b)
=-2{\bf k}R_u(u,v;a,b);\\
V^{(a)}(k,p)&=&\int dy~y_4D(y)W(|{\bf y}|,|y_4|)e^{-iky}\\
&=&-2k_4{\partial\over\partial v}\tilde{R}(u,v;a,b)
=-2k_4R_v(u,v;a,b),
\end{eqnarray*}
where the function $R$ is defined by (\ref{R}), and
\begin{eqnarray*}
({\bf \Psi}^{(n)}D{\bf \Psi}^{(n)})&=&\int{dy~{\bf y}^2\over(2\pi)^2y^2}
e^{-2a\sqrt{{\bf y}^2+by_4^2}}
={3\over16a^4}\cdot{2+\sqrt{b}\over(1+\sqrt{b})^2},\\
(\Psi^{(a)}D\Psi^{(a)})&=&\int{dy~y_4^2\over(2\pi)^2y^2}
e^{-2a\sqrt{{\bf y}^2+by_4^2}}
={3\over16a^4}\cdot{1\over\sqrt{b}(1+\sqrt{b})^2},\\
\end{eqnarray*}
The eigenvalue $\Lambda(M^2)$ is defined by
\begin{eqnarray}
\label{varmassvs}
&& g^2\Lambda^{(n)}(M)=\alpha\max\limits_{a,b}{8\over\pi^2}
\int\limits_0^\infty dk~k^2\int\limits_0^\infty dv\cdot
{16a^4(1+\sqrt{b})^2 4k^2R_u^2(k^2,v^2;a,b)\over3(2+\sqrt{b})
\left(k^2+v^2+1-{M^2\over4}\right)^2+M^2v^2},\nonumber\\
&& g^2\Lambda^{(a)}(M)=\alpha\max\limits_{a,b}{8\over\pi^2}
\int\limits_0^\infty dk~k^2\int\limits_0^\infty dv\cdot
{16a^4\sqrt{b}(1+\sqrt{b})^2 4q^2R_v^2(k^2,v^2;a,b)\over
3\left(k^2+v^2+1-{M^2\over4}\right)^2+M^2v^2}.
\end{eqnarray}
The value of $M$ is defined by equation (\ref{mass}).
The nonrelativistic binding energy of the bound state with $l=1$ is
\begin{eqnarray*}
&& \epsilon=2-M={\alpha^2\over16},~~~{\rm and}~~~
{1\over\alpha}={1\over\sqrt{32\left(1-{M\over2}\right)}}=
{1\over\alpha_{Sch}},~~~~~\alpha_{em}={1\over137}
\end{eqnarray*}
The numerical results are represented in Table 2 where
$a=\sqrt{\Delta}a_1$ and $b=\Delta b_1$. Here we should remark that
the calculations in \cite{Tjon} contain a mistake for $l\geq1$.
For the state with $l=1$ factor ${1\over2}$ should be introduced to
get the correct value.

\begin{center}
{Table 2. The first excited states in the Wick-Cutkosky model. The coupling
constant $\lambda=4\alpha$ for different masses $M/m$ and comparison
$1/\alpha$ with the nonrelativistic value $1/\alpha_{Sch}$.}
\end{center}
\begin{center}
\begin{tabular}{|c||c|c||c||c|c||c|r||r|l|r|}
\hline
\multicolumn{3}{|c||}{} & \multicolumn{5}{|c||}{} & \multicolumn{3}{|c|}{}\\
\multicolumn{3}{|c||}{} & \multicolumn{5}{|c||}{Vector particles $l=1$}
& \multicolumn{3}{|c|}{"Abnormal state"}\\
\multicolumn{3}{|c||}{} & \multicolumn{5}{|c||}{} & \multicolumn{3}{|c|}{}\\
\hline
& & & & & & & & & &\\
${M\over m}$&${\epsilon\over2m}$&${1\over\alpha_{Sch}}$&${1\over\alpha}$&
$a_1$&$b_1$&$\lambda_v=4\alpha$&$\lambda_{Tj}$&$\lambda_s$&$a_1$&$b_1$ \\
& & & & & & & & & &\\
\hline
& & & & & & & & & &\\
0.000 &1.00 &0.176 &.0529 &0.80 &1.00 &75.62 & 80 &75.55 &0.85 &1.0\\
0.500 & 0.75 & 0.204  & 0.0559 & 0.85 &1.00 &71.47 & &72.78 &0.85 &1.0  \\
1.000 & 0.50 & 0.250  & 0.0676 & 0.81 & 1.20 &59.19 &63.5 &64.27 &0.90 &1.1\\
1.500 & 0.25 & 0.353  & 0.1059 & 0.87 & 1.48 & 37.77 & &48.55 &0.94 &1.4 \\
1.900 & 0.05 & 0.790  & 0.3409 & 0.89 &2.56 &11.73 &16 &25.63 &1.05 &2.85\\
1.999 & 0.0005 & 7.906& 6.4894 & 0.95 & 5.00 & 0.616 & 5 &8.98 &1.30 &14.0\\
$2-{\alpha_{em}^2\over16}$ & ${\alpha_{em}^2\over32}$
& 137 &133.9 & 0.986 & 10.0 & 0.0298 & & & & \\
& & & & & & & & & &\\
\hline
\end{tabular}
\end{center}

One can see that "abnormal states" have practically the same mass as
"normal" states in the strong coupling regime when the mass defect is large.
If the coupling constant is small the "abnormal" masses are much more
then "normal" masses. In addition we can repeat that the nonrelativistic
regime is realized for very small binding energies.

\section{Massive tieons $\mu>0$}

In this section we perform variational calculations in the case $\mu>0$
for the lowest and the first excited states and compare our numerical
results with the numbers obtained in \cite{Tjon}.

\subsection{The lowest state}

For the lowest state we choose the test function in the form
\begin{eqnarray}
\label{Psitest00}
&& \Psi(y,p)=D_3(y,a,b)=
\int{dk\over(2\pi)^4}\cdot{e^{iky}\over(a^2+{\bf k}^2+bk_4^2)^3}
\end{eqnarray}
Here $a$ and $b$ are variational parameters.

The direct calculations give
\begin{eqnarray}
\label{Zn}
&& (\Psi D\Psi)=\int dy~D(y)D_3^2(y,a,b)={1\over2(4\pi)^4}\cdot
Z(\mu,a,b),\\
&& Z(\mu,a,b)=
\int\!\!\!\int\limits_0^1{dtdu~u^3t(1-t)\over\sqrt{b(bu+1-u)}~
(a^2u+\mu^2(1-u)t(1-t))^3},\nonumber
\end{eqnarray}
\begin{eqnarray}
\label{Ch}
&&\int dy~D(y)\Psi(y)e^{iky}=\int dy D(y)D_3(y,a)e^{iky}=
{1\over2(4\pi)^4}\cdot V(r,\theta,a,b),\\
&& V(r,\theta,a,b)=\int\limits_0^1{dt~t^2\over\sqrt{bt+1-t}}\cdot
{1\over\left[a^2t+\mu^2(1-t)+t(1-t)r^2
\left(\cos^2(\theta)+{\sin^2(\theta)b\over bt+1-t}\right)\right]^2},
\nonumber
\end{eqnarray}
where we put $|{\bf k}|^2=r^2\cos^2(\theta)$ and $k_4^2=r^2\sin^2(\theta)$.

The mass of the lowest state is defined by the variational equation
(\ref{VPsi}). Our formulation of the problem is to find the coupling
constant $\lambda=4\alpha$ for given $\mu={\mu\over m}$ and
$M={M\over m}$, then we get
\begin{eqnarray}
\label{lambda00}
\lambda&=&\min\limits_{a,b}{\pi^2Z(\mu,a,b)\over J(\mu,M,a,b)}
\end{eqnarray}
where
\begin{eqnarray*}
J(\mu,M,a,b)&=&\int\limits_0^\infty dr\int\limits_0^{{\pi\over2}}d\theta
{r^3\cos^2(\theta)~V^2(r,\theta,a,b)\over
\left(r^2+1-{M^2\over4}\right)^2+M^2r^2\sin^2(\theta)}
\end{eqnarray*}
The numerical results are listed in Table3.

\begin{center}
{Table 3. The lowest states in the case $\mu>0$.}
\end{center}
\begin{center}
\begin{tabular}{|c|c||c|c||c|c|c||c|}
\hline
& & & & & & & \\
${\mu\over m}$ & ${M\over m}$ & $\lambda=4\alpha$ & $a$ &
$\lambda=4\alpha$ & $a$ & $b$ & $\lambda_{Tj}$ \\
&  &  & $b=1$ &  &  & &  \\
& & & & & & & \\
\hline
& & & & & & & \\
1.000 & 0.000 & 42.960 & 1.27  & 42.960 & 1.27 & 1.00 & 42.96\\
1.000 & 1.000 & 36.945 & 1.15  & 36.940 & 1.27 & 1.10 & 36.94\\
1.000 & 1.900 & 17.293 & 0.66  & 17.252 & 0.70 & 1.87 & 17.23\\
1.000 & 1.999 & 10.294 & 0.41  & 10.273 & 0.44 & 2.35 & 10.25\\
& & & & & & & \\
0.100 & 0.000 & 25.802 & 1.02  & 25.802 & 1.02 & 1.00 & 25.80\\
0.100 & 1.000 & 20.692 & 0.89  & 20.680 & 0.91 & 1.20 & 20.68 \\
0.100 & 1.900 & 5.3843 & 3.50  & 5.262  & 0.38 & 2.22 & 5.227 \\
0.100 & 1.999 & 1.0789 & 0.089  & 1.054 & 0.089 & 9.00 & 1.043\\
& & & & & & & \\
0.001 & 0.000 & 25.133 & 1.00  & 25.133 & 1.00 & 1.00 & 25.13\\
0.001 & 1.000 & 20.023 & 0.87  & 20.013 & 0.87 & 1.20 & 20.01\\
0.001 & 1.900 & 4.6904 & 0.32  & 4.501  & 0.35  & 4.20 & 4.483\\
0.001 & 1.999 & 0.3523 & 0.031 & 0.2900 & 0.04  & 150  & 0.289\\
& & & & & & & \\
\hline
\end{tabular}
\end{center}

In Table 4 we listed the results of calculations for two cases:
(1) one parameter $a$ with $b=1$ and (2) two parameters $a$ and $b$.
One can see that only for small binding energies there is a remarkable
difference in numbers. Our numbers are quite close to the numerical
results obtained in \cite{Tjon}.

\subsection{The first excited states}

For the first excited states we choose the test functions for
"normal" $(n)$ and "abnormal" $(a)$ states in the form
\begin{eqnarray}
\label{Psitest11}
&& \Psi_\mu(y,p)=y_\mu D_3(y,a,b)=
\left\{\begin{array}{l}
\vec{\Psi}^{(n)}(y,p)={\bf y}D_3(y,a,b),\\
\Psi^{(a})(y,p)=y_4 D_3(y,a,b)
\end{array}\right.
\end{eqnarray}

Repeating as before all calculations we reduce the problem to find
the coupling constant $\lambda=4\alpha$ for given $\mu={\mu\over m}$ and
$M={M\over m}$ to the variational task
\begin{eqnarray}
\label{lambda11}
\lambda&=&\min\limits_{a,b}{\pi^2Z_j(\mu,a,b)\over J_j(\mu,M,a,b)},
~~~~(j=n,a).
\end{eqnarray}
Here
\begin{eqnarray*}
&& Z_n(\mu,a,b)=
\int\!\!\!\int\limits_0^1{dtdu~18u^4(1-u)t^2(1-t)^2\over\sqrt{b(bu+1-u)}
(a^2u+\mu^2(1-u)t(1-t))^4},\\
&& Z_a(\mu,a,b)=
\int\!\!\!\int\limits_0^1{dtdu~6u^4(1-u)t^2(1-t)^2\over\sqrt{b(bu+1-u)^3}
(a^2u+\mu^2(1-u)t(1-t))^4},
\end{eqnarray*}

\begin{eqnarray*}
J_j(\mu,M,a,b)&=&\int\limits_0^\infty dr\int\limits_0^{{\pi\over2}}d\theta
{r^3\cos^2(\theta)V_j^2(r,\theta,a,b)\over
\left(r^2+1-{M^2\over4}\right)^2+M^2r^2\sin^2(\theta)}
\end{eqnarray*}

\begin{eqnarray*}
&& V_n(r,\theta,a,b)=\int\limits_0^1{dt~t^3\over\sqrt{bt+1-t}}\cdot
{4r\cos(\theta)\over h(\mu,r,\theta,t,a,b)},\\
&& V_a(r,\theta,a,b)=\int\limits_0^1{dt~t^3\sqrt{b}\over\sqrt{(bt+1-t)^3}}
\cdot{4r\sin(\theta)\over h(\mu,r,\theta,t,a,b)},
\end{eqnarray*}
$$ h(\mu,r,\theta,t,a,b)=\left[a^2t+\mu^2(1-t)+t(1-t)r^2
\left(\cos^2(\theta)+{\sin^2(\theta)b\over bt+1-t}\right)\right]^3.$$

In Table 4 the results of calculations for "normal" states
with spin $l=1$ and for "abnormal" states with spin $l=0$ are listed.
The column $\lambda_{Tj}$ contains the results obtained in \cite{Tjon}
with factor ${1\over2}$ to get the correct numbers. One can see that our
numbers are quite close to  them. It should be noted that in the strong
coupling regime the "abnormal" states begin to play an essential role
in the representation of the BS amplitude (\ref{Bst}).

\vspace{1cm}

{\it Acknowledgments}. I would like to thank S.Dorkin for useful
discussions and help in numerical calculations. This work was
supported by the Russian Foundation for Basic Research under
grant No. 01-02-17200-a.

\begin{center}
{Table 4. The first excited states in the case $\mu>0$.}
\end{center}
\begin{center}
\begin{tabular}{|c|c||r|c|c|r||r|l|r|}
\hline
\multicolumn{2}{|c||}{} & \multicolumn{4}{|c||}{} & \multicolumn{3}{|c|}{}\\
\multicolumn{2}{|c||}{} & \multicolumn{4}{|c||}{Vector particles $l=1$}
& \multicolumn{3}{|c|}{"Abnormal state"}\\
\multicolumn{2}{|c||}{} & \multicolumn{4}{|c||}{} & \multicolumn{3}{|c|}{}\\
\hline
& & & & & & & & \\
${\mu\over m}$ & ${M\over m}$ & $\lambda_n=4\alpha$ & $a$ & $b$
& $\lambda_{Tj}$&$\lambda_s=4\alpha$ & $a$ & $b$  \\
& & & & & & & & \\
\hline
& & & & & & & &  \\
1.000 & 0.000 & 205.95 & 1.6  & 1.0 &   & 205.95 & 1.6 & 1.0\\
1.000 & 1.000 & 183.16 & 1.5  & 1.1 &   & 195.35 & 1.5 & 1.1\\
1.000 & 1.900 & 110.86 & 1.1  & 1.5 &   & 166.26 & 1.3 & 1.3\\
1.000 & 1.999 & 93.48 &  0.9  & 1.9 &   & 161.66 & 1.3 & 1.5\\
& & & & & & & & \\
0.100 & 0.000 & 80.35 & 1.1  & 1.1  & 80.22 & 80.30 & 1.0 & 1.0\\
0.100 & 1.000 & 63.77 & 0.9  & 1.2  & 63.77 & 63.74 & 0.9 & 1.1 \\
0.100 & 1.900 & 16.05 & 0.4  & 3.2  & 15.99 & 34.78 & 0.6 & 3.2 \\
0.100 & 1.999 & 4.87 & 0.2  & 6.2  & 4.77  & 27.44 & 0.4 & 4.4\\
& & & & & & & & \\
0.001 & 0.000 & 75.40 & 1.0 & 1.1  &     & 75.40 & 1.0 & 1.0 \\
0.001 & 1.000 & 59.06 & 0.9  & 1.2 &     & 64.24  & 0.9 & 1.2\\
0.001 & 1.900 & 11.84 & 0.3  & 3.2 &     & 25.75 & 0.4 & 4.1\\
0.001 & 1.999 & 1.03  & 0.1  & 5.2 &     & 10.11  & 0.05  & 15.0\\
& & & & & & & & \\
\hline
\end{tabular}
\end{center}

\section{Appendix. Angular polynomials $T^{(l)}_{\{\mu\}}(k)$}

The angular polynomials $T^{(l)}_{\{\mu\}}(k)=T^{(l)}_{\mu_1,...,\mu_l}(k)$
are symmetric for $\mu_i\rightleftharpoons\mu_j$, and
$$T^{(l)}_{\mu,\mu,\mu_3,...,\mu_l}(k)=0$$
The recurrence relation is
$$T^{(l)}_{1...l}(k)={1\over l}P(1\vert2...l)k_1T^{(l-1)}_{2...l}(k)
-{k^2\over2l(l-1)}P(12\vert3...l)\delta_{12}T^{(l-2)}_{3...l}(k)$$
In particular
\begin{eqnarray*}
&& T^{(0)}=1,~~~~~~~~~~T^{(1)}_1(k)=k_1,\\ &&
T^{(2)}_{12}(k)=k_1k_2-{1\over4}\delta_{12}k^2,\\ &&
T^{(3)}_{123}(k)=k_1k_2k_3-{k^2\over6}(k_1\delta_{23}+
k_2\delta_{31}+k_3\delta_{12}),\\
&& .....................................
\end{eqnarray*}
The normalization condition is
\begin{eqnarray*}
&& \sum\limits_{\mu_1...\mu_l}T^{(l)}_{\mu_1,...,\mu_l}(k)
T^{(l)}_{\mu_1,...,\mu_l}(p)={|k|^l|p|^l\over2^l}C^1_l((n_kn_p)),
\end{eqnarray*}
where $C^1_l(t)$ is the Gegenbauer polynomial.

These angular functions satisfy the formulas
\begin{eqnarray}
\label{Tl}
&& \Box_kT^{(l)}_{\mu_1,...,\mu_l}(k)=0,~~~~~
\left(k{\partial\over\partial k}\right)T^{(l)}_{\mu_1,...,\mu_l}(k)=
l~T^{(l)}_{\mu_1,...,\mu_l}(k).
\end{eqnarray}


\begin{thebibliography}{99}
\bibitem{Wick}
G.C.Wick, Phys.Rev., {\bf 96}, 1124 (1954); R.E.Cutkosky,
Phys.Rev., {\bf 96}, 1135 (1954).
\bibitem{Itz}
C.Itzykson and J.-B.Zuber, {\sl Quantum Field Theory}, McGraw-Hill
Book Company, N.Y., 1980.
\bibitem{Sapir}
J.R.Sapirstein and D.R.Yennie, {\sl Theory of Hydrogenic Bound
States} in {\sl Quantum Electrodynamics}, Ed. T.Kinishita,
Advanced Series on Directions in High Energy Physics, vol 7, World
Scientific, 1999.
\bibitem{Grein}
W.Greiner and J.Reinhart, {\sl Quantum Electrodinamics},
Springer-Verlag, Berlin, N.Y., 1992.
\bibitem{Nak}
N.Nakanishi, Suppl. of Pr.Theor.Phys., {\bf 43}, 1 (1969).
\bibitem{Tjon}
T.Nieuwenhuis, J.A.Tjon, Few Body Systems, {\bf 21}, 167 (1996).
\bibitem{Kapt}
S.M.Dorkin, L.P.Kapteri and S.S.Semikh, Russ.Jad.Fiz, {\bf 60}, 1784 (1997).
\bibitem{Vosko}
S.H.Vosko, J.Math.Phys., {\bf 1}, 505 (1960).
\bibitem{Schw}
C.Schwarz, Phys.Rev., {\bf 137B}, 717 (1965).
\bibitem{Simon}
B.Simon, {\sl The $P(\phi)_2$ Euclidian (Quantum) Field Theory},
Princeton Univ. Press, Princeton, 1974.
\bibitem{Bog}
N.N.Bogoliubov, D.V.Shirkov, {\sl Introduction to the Theory of
Quantized Fields}, Interscience Publishers, Ltd, N.Y., 1959.
\bibitem{Dar}
J.W.Darewych, Can.J.Phys., {\bf 76}, 523 (1998).
\end{thebibliography}
\end{document}